\documentclass[prb]{revtex4}
\usepackage{graphicx}
\usepackage{dcolumn}
\usepackage{bm}
\usepackage{amsmath}

\begin{document}

\preprint{APS/123-QED}

\title{Magic Angle Electron Energy Loss Spectroscopy (MAEELS) \\ of core electron excitation in anisotropic systems}

\author{Y.K.Sun }
\author{J.Yuan}%
\email{yuanjun@mail.tsinghua.edu.cn}
\affiliation{%
Department Materials Science and Engineering, Tsinghua University,
China
}%

\date{\today}

\begin{abstract}
A general theory for the core-level electron excitation of
anisotropic systems using angular integrated electron energy-loss
spectroscopy has been derived.  We show that it is possible to
define a magic angle condition at which the specimen orientation
has no effect on the electron energy-loss spectra.  We have not
only resolved the existing discrepancy between different studies
of the magic angle condition, but also extended its applicability
to all anisotropic systems.  We have demonstrated that magic angle
electron energy loss spectroscopy is equivalent to the orientation
averaged EELS, although the specimen remains stationary.  Our
analysis provides the theoretical framework for the comparison
between theoretical calculation and experimental measurement of
core-level electron excitation spectra in anisotropic systems. In
addition to MAEELS, we have also discovered a magic orientation
condition which will also give rise to orientationally averaged
spectra.  It's relation with the magic angle X-ray absorption
spectroscopy and magic angle spinning nuclear magnetic resonance
is demonstrated.

\end{abstract}
\pacs{Valid PACS appear here}
\maketitle

\section{\label{sec:level1}Introduction}

Anisotropic systems differ from isotropic ones in that their
response to an applied force or field depends not only on the
magnitude but also the orientation of these
influences\cite{JUR74}. For electron energy loss spectroscopy
(EELS) or X-ray absorption spectroscopy (XAS), a good example is
the carbon 1s core electron excitation in
graphite\cite{LEA79,LEA83,DIS82,BAS93,KIN88,STH92}.  Excitation
into the unoccupied states of $\pi$-symmetry is only allowed if
the applied field is along the direction normal to the graphite
sheet (defined to be the local z-axis). Excitation into the
unoccupied state of $\sigma$-symmetry can only occur if the
applied field lies in the plane of the graphite sheet (defined to
be the local x-y plane). As a consequence, the intensities of
these two excitations depend on the specimen orientation in
general. Graphite is an uniaxial system which is the simplest
example of anisotropic systems.  Anisotropic response is a gift to
the experimentalists as it offers further insight into the
electronic excitation process\cite{KIN88,STH92,EGE96} or it can be
used to determine the orientation of molecules\cite{RUI97,VAN00}
or internal magnetic field\cite{RAH86}.  In this paper, we focus
on the complexity it brings to the EELS measurement which now also
depends on the precise orientation of the specimen and ways to
overcome it.

Concerning electronic excitation, many important systems show
anisotropy such as familiar layered materials like graphite and
BN\cite{LEE00}, non-central symmetric semiconducting compounds
such as GaN\cite{LUB99}, superconductors such as
YBaCuO\cite{ZHU93,ZHU94,FIN94,NUC95} and MgB$_2$
\cite{ZHU02,KON02,ZHA03,KLI03}. Some nanostructures such as
nanotubes and nanoonions can be considered as roll-up versions of
layered materials, so the local anisotropy also changes with
position inside the nanostructured material\cite{SOU98}. In
addition, shape or local field effects will turn an isotropic
transition into an anisotropic one\cite{VAS02}. This presents
difficulty for quantitative spectral analysis, particularly from
localized areas using EELS, a powerful method for electronic
structural characterization, particularly at nanoscale.  For
example, real changes in the electronic structure may not be
easily distinguished from variations due to mere specimen rotation
or probe displacement along a curved basal plane. So there is a
need to find an experimental approach in which specimen
orientation is no longer effective on the spectra.

A related problem also exists in other forms of spectroscopy.  For
instance, in solid state nuclear magnetic resonance (NMR), the
signal depends strongly on the specimen orientation and more
useful information is obtained after the discovery of the magic
angle spinning nuclear magnetic resonance (MAS-NMR)
method\cite{RAH86,AND59,LOW59}. Another example in XAS is the
so-called 'magic angle' condition used to eliminate the
orientation effect for anisotropic systems\cite{STH92,KIN88}. In
EELS, the applied field responsible for electronic excitation is
parallel to the direction of the momentum transfer vector
\textbf{q} of the incident electron in the inelastic
scattering\cite{EGE96}. As shown in fig.\ref{fig1}, by virtue of
electron scattering, the direction of \textbf{q} is a function of
the electron scattering angle $\theta$, being parallel to the
incident electron beam at the zero scattering angle and changing
towards a direction perpendicular to the beam with the increasing
scattering angle.

In the angular integrated energy-loss spectroscopy with a centered
circular aperture, a standard EELS collection condition
particularly for high spatial resolution and high signal to noise
ratio, electronic excitations of different momentum transfer
\textbf{q} are recorded. Excitations induced by the applied fields
over a range of directions are
summed\cite{BRO91,BRO93,BOT95,MEN98}. Menon and Yuan\cite{MEN98}
showed both experimentally and theoretically there also exists a
magic angle (MA) at which the EELS signal is independent of
specimen orientation. Their theory predicts the magic angle for a
parallel beam illumination (MA$_{\parallel}$) to be $4\theta_{E}$.
There $\theta_{E}$ is the characteristic angle which has a
approximate form ${\theta}$$_{E}$=2E/E$_{0}$ (the more accurate
definition is given by Ritchie and Howie\cite{RIT77}) and with E
being the energy-loss and E$_{0}$ the incident energy of the fast
electrons. The derivation is for uniaxial, graphite-like, systems.

A number of other studies have also analyzed this problem again,
mostly for uniaxial anisotropic systems. However, controversy as
to the precise magic angle condition persists. For example, Zhu et
al.\cite{ZHU93} used an approximate method to arrive at a value of
MA$_{\parallel}$ of about 1.8${\theta}$$_{E}$.  Paxton et
al.\cite{PAX00} used a different approach to arrive at
MA$_{\parallel}$=1.36 ${\theta}$$_{E}$. Compounding this state of
confusion in magic angle prediction is the report of Daniels et
al.\cite{DAN03} who found experimentally that
MA$_{\parallel}$=2${\theta}$$_{E}$ and also provided a theoretical
justification for their result. Souche et al.\cite{SOU98}
presented a more through theoretical analysis of the anisotropic
electron energy-loss spectroscopy for graphite-like materials,
taking into account the convergence beam effect.  Their result for
the parallel beam illumination is in agreement with the prediction
of Menon and Yuan\cite{MEN98}. In addition, apart from attempts by
Paxton, all theoretical analysis has been confined to MAEELS in
uniaxial systems, the applicability of the magic angle concept in
more complex anisotropic systems is an unexplored area.

In order to make use of MAEELS, it is vital to understand the
discrepancies between the various theories, particularly given
that all models make the same basic assumptions, i.e. the electron
beam scattering being kinetic, non-relativistic, and obeying the
dipole approximation.  In the process, we have demonstrated that
the magic angle is a general effect which can occur in all sorts
of anisotropic systems, be they crystalline, amorphous, single
phase or nanostructured and applies to anisotropic transition
whether it is intrinsic or extrinsic (for example, shape-effect
induced).

The paper is organized as follows. In section II, we present a
through analysis of electron energy-loss spectroscopy in
anisotropic systems in order to establish a solid foundation for
MAEELS in the core-electron excitation region and we derive a
general definition for magic angle conditions. In section III, we
show that the spectral information obtained at the magic angle is
equivalent to orientationally averaged spectroscopy.  From this
geometrical interpretation, we can understand the different
approaches used in MAEELS analysis reported in the literature and
explain the reasons for discrepancies in the prediction of magic
angle conditions. In section IV, we have provided more explicit
MAEELS expressions for applications in specific material systems.
In particular, we concentrated on more commonly encountered
uniaxial and biaxial systems and establish the connection between
MAEELS and other forms of magic angle spectroscopies such as XAS
and MAS-NMR.

\section{Generalized MAEELS Theory}
There are two ways to study the electronic excitation in
anisotropic systems: the macroscopic approach is to use the
dielectric function and the microscopic approach to calculate
directly the quantum mechanical transition matrix element. Both
approaches have been reported in the
literature\cite{ZHU93,SOU98,BRO91,BRO93,PAX00,DAN03,GLO00} and
physically they should yield the same result. To facilitate
comparison of these reports, we will make our derivation using
both approaches. It is known that the quantum mechanical approach
can reveal the microscopic physics involved, but because of the
need for accurate wave functions, it does not always yield useful
experimental results. On the other hand, the dielectric approach
is a phenomenological description of materials response that can
be measured accurately, even though the microscopic origin of the
electronic transitions may be obscured\cite{VAS02}.

\subsection{\label{sec:level2}Dielectric Formalism}

Here we start with the dielectric approach where the calculation
is relatively straightforward, because the information required is
not the detailed excitation form of but the overall effect in
terms of the response to a perturbation deserved by the
relation\cite{JAC99}:
\begin{eqnarray}
D_{i}=\sum_j \varepsilon_0\varepsilon^{ij}E_j
\end{eqnarray}
where \textbf{D}, the electric displacement, is related to the
electric field \textbf{E} by the well-known $\varepsilon^{ij}$ the
dielectric function of the material system. $\varepsilon^{ij}$ is
a 'metric tensor'\cite{PRI94}, defined in terms of a reference
frame where the orthogonal principle axes are aligned with the
major symmetry directions of the physical system. For convenience,
we have defined this reference frame as the sample frame (x,y,z).

\begin{figure}
\begin{center}
  \includegraphics{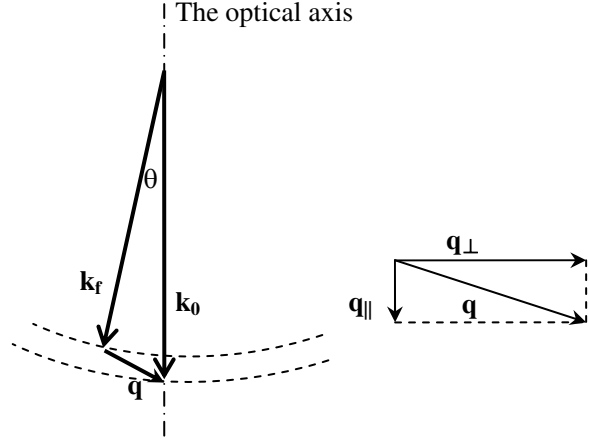}
  \caption{In an inelastic scattering, the momentum transfer vector \textbf{q}
   is determined by the initial and final wave-vector \textbf{k}$_0$ and \textbf{k}$_f$.
   The angle between the two wave vectors is defined as the scattering angle $\theta$. In the parallel illumination case, $\theta$ always equals
   to $\beta$ defined as the angle between the wave vector of scattering electrons $k_f$ and the optical axis.}\label{fig1}
\end{center}
\end{figure}

The imaginary parts of (-1/$\varepsilon$) is known as the
energy-loss function\cite{EGE96,BRO91,BRO93,MEN98}, providing a
complete description of the response of the medium through which
the fast electron is travelling. The double differential cross
section used to estimate the intensity of EELS in an anisotropic
system can be expressed as\cite{EGE96,BRO93}:
\begin{eqnarray}
\frac{d\sigma^2(\textbf{q})}{dEd\Omega}=\frac{4m_e}{na_0h^2}
\textbf{\emph{Im}}(-\frac{1}{{\displaystyle\sum_{i,j}} q_{i}
\varepsilon^{ij}q_{j}})\label{a2}
\end{eqnarray}
where \emph{m$_e$} the mass of the fast electron and \textbf{q} is
the momentum transfer vector of the fast electrons in the
inelastic scattering process (see fig.\ref{fig1}),  \emph{n} the
number of atoms per unit volume of the material, \emph{a$_0$} the
Bohr atomic radius, \emph{h} the Plank constant and q$_i$ the
projection of \textbf{q} in the sample frame. Note the components
of the dielectric function are assumed not to be a function of
\textbf{q}, and this assumption corresponds to the dipole
approximation in the quantum mechanical analysis of single
electron transitions\cite{JAC99}.

We will restrict our discussion to core electron excitations in
which we use the approximation\cite{EGE96,BRO91,BRO93} for
$\varepsilon_1$=Re($\varepsilon$) and
$\varepsilon$$_2$=Im($\varepsilon$), $\varepsilon$$_{1}$ $\approx
1$ and $\varepsilon$$_{2}$ $\approx 0$.  Eq.(\ref{a2}) can then be
simplified to
\begin{eqnarray}
\frac{d\sigma^2(\textbf{q})}{dEd\Omega}=\frac{4m_e}{na_0h^2}
\sum_{i,j} \frac{q_{i} \varepsilon{_2}^{ij}q_{j}}{q^4}\label{a3}
\end{eqnarray}

\begin{figure}
  \includegraphics{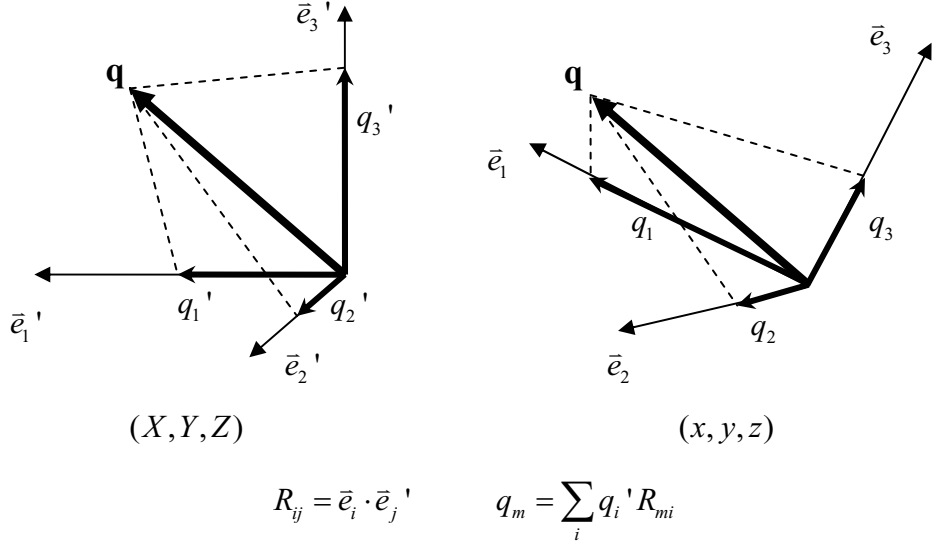}
  \caption{The projection of the momentum transfer vector \textbf{q} in the sample frames (x,y,z) and in the laboratory frame
  (X,Y,Z). The components $q_i$ ($q_i$') are equivalent to $\textbf{q}$$\cdot$$\textbf{e}_i$ ($\textbf{q}$$\cdot$$\textbf{e}_i$'),
    where the $\textbf{e}_i$ ($\textbf{e}_i$') are the basis vectors of the reference frame.  Two sets of
  components of \textbf{q} are related by the transformation matrix R whose elements are defined in terms of dot product of the basis vectors of the two reference frames.}\label{fig2}
\end{figure}

From now on, we will be only interested in the EELS which is
obtained by integrating Eq.(\ref{a3}) over the angular range
determined by the collection condition, i.e. the convergence
semi-angle $\alpha_0$ for a convergent beam and the collection
semi-angle $\beta_0$ for the centered collection aperture used.
This gives the partial angular integrated cross section
\begin{eqnarray}
\frac{d\sigma}{dE}(\alpha_0,\beta_0,\widetilde{O})
=\frac{4m_e}{na_0h^2}\sum_{i,j}(\int d\Omega
\frac{q_iq_j}{q^4})\varepsilon{_2}^{ij} =\frac{8\pi
m_e}{na_0h^2k_0^2}\sum_{i,j}W_{ij}(\alpha_0,\beta_0,\widetilde{O})\varepsilon{_2}^{ij}\label{a4}
\end{eqnarray}
where $\widetilde{O}$ denotes the orientation of the sample with
respect to the electron beam direction. Here the weighting factor
W$_{ij}$ depends both on the specimen orientation and  the
experimental condition used, i.e. $\alpha_0$ and $\beta_0$.  In
order to separate out these two effects, we have transformed the
representation of \textbf{q} from the (x,y,z) orthogonal
coordinate of the sample frame to the (X,Y,Z) orthogonal
coordinate of the laboratory frame, with the Z axis defined to be
optical axis of the electron beam.  We denote the components of
\textbf{q} in the (X,Y,Z) frame as q'$_{i}$(fig.\ref{fig2}). Two
representations of \textbf{q} are related to each other through
the rotational transformation matrix \textbf{R} as:
\begin{eqnarray} q_m=\sum_i q'_{i}R_{mi}\label{a5}
\end{eqnarray}
For transformation between two orthogonal coordinate systems, the
rotational matrix element R$_{ij}$ is defined to be the direction
cosine between the basis vector \textbf{e}\textbf{'} in the
(X,Y,Z) frame and the basis vector \textbf{e} in the (x, y, z)
frame\cite{JUR74,PRI94}:
\begin{eqnarray} R_{ij}=\textbf{e}_i \cdot \textbf{e}_j'\label{a6}
\end{eqnarray}
Substituting Eq.(\ref{a5}) into Eq.(\ref{a4}), we get the
definition for the weighting factor as:
\begin{eqnarray} W_{ij}(\alpha_0,\beta_0,\widetilde{O})=\frac{k_0^2}{2\pi}\sum_{m,n}(\int d\Omega
\frac{q_m'q_n'}{q^4})R_{mi}R_{nj}\label{a7}
\end{eqnarray}
In this way we have successfully separated out the orientation
factors, in the form of the product of matrix elements, from the
integral within the bracket which is solely determined by the
experimental set-up. By inspection, we can see that the
integration over the full azimuthal angle of vector \textbf{q} of
the integrand with the cross-indices vanishes because of the
rotational symmetry.  This means that the integral has the
simplified forms as follows:
\begin{eqnarray}
\int d\Omega \frac{q_m'q_n'}{q^4}= \left\{
\begin{array}{l@{}l}
\displaystyle\int d\Omega \frac{q_{\parallel}^2}{q^4}=\frac{2\pi}{k_0^2}\xi_{\parallel}(\alpha_0,\beta_0) & (m=n=3)\\
\displaystyle\frac {1}{2}\int d\Omega \frac{q_{\perp}^2}{q^4}=
\frac{2\pi}{k_0^2}\xi_{\perp}(\alpha_0,\beta_0)& (m=n=1 or 2)\\
\displaystyle 0 & (m\neq n) \label{a8}
\end{array}
\right.
\end{eqnarray}
where we have introduced notation q$_{\parallel}$ and q$_{\perp}$
to denote the components of \textbf{q} that are parallel and
perpendicular to the incident beam direction respectively
(fig.\ref{fig1}) and k$_0$ is the magnitude of the wave-vector for
the fast electron beam.  The factor
$\displaystyle\frac{2\pi}{k_0^2}$ is used to eliminate the
dimension of the reduced integral variable $\xi_{\parallel}$ and
$\xi_{\perp}$.

Putting the integral expression in Eq.(\ref{a8}) back into
Eq.(\ref{a7}) the weighting factor can be written as:
\begin{eqnarray}
W_{ij}(\alpha_0,\beta_0,\widetilde{O})=
\xi_{\parallel}R_{3i}R_{3j}+\xi_{\perp}(R_{1i}R_{1j}+R_{2i}R_{2j})
\end{eqnarray}

If we rearrange the product of the matrix element by applying the
orthogonal property of the transformation matrix\cite{JUR74} as
$\sum\limits_m R_{mi}R_{mj}=\delta_{ij}$, and through explicit
calculation using Eq.(\ref{a6}), we can obtain a more revealing
definition for the weighting factor
\begin{eqnarray}
W_{ij}(\alpha_0,\beta_0,\widetilde{O})=\xi_{\perp}\delta_{ij}+(\xi_{\parallel}-\xi_{\perp})\cos\chi_{i}\cos\chi_j\label{a10}
\end{eqnarray}
where the $\chi_i$ is the angle between the optical axis (the
Z-direction) of the laboratory frame and the \emph{ith} basis
vector in the sample frame.  It is clear that the second term in
Eq.(\ref{a10}) gives the information about orientation of the
sample, so the magic angle condition is defined by the relation
\begin{eqnarray}
\xi_{\parallel}(\alpha_{0}^{MA},\beta_{0}^{MA})=\xi_{\perp}(\alpha_{0}^{MA},\beta_{0}^{MA})\label{macondition}
\end{eqnarray}

Note that our derivation has not exploited any special symmetry
properties of the dielectric function, for example, some specific
symmetry property brought by a crystal structure. Hence the magic
angle condition is valid for all anisotropic systems, i.e., it not
only applies to single crystals, but also to amorphous materials,
powders, or nanostructures as long as an effective dielectric
function tensor can be defined and measured.

\subsection{Quantum Mechanical Theory}
The dielectric function approach is very useful in treating
practical problems. However, to understand the microscopic origin
of the electronic transition responsible, it is better to work
explicitly in terms of a quantum mechanical theory. With the wide
spread use of \emph{ab-initio} quantum mechanical calculation
methods, proper treatment of the electronic excitation will become
be routine\cite{NEL99,HEB00,SCH01}. It is vital to know how to
relate them to measurements where specimen-orientation is an
additional variable.

IN quantum mechanics the inelastic scattering of high energy
electrons can be described adequately by the first Born
approximation as\cite{BET64,INO71}:
\begin{eqnarray}
\frac{d\sigma^2(\textbf{q})}{dEd\Omega}=\frac{4}{a_0^2}\cdot\frac{1}{q^4}|\langle
f|\exp(-i\textbf{q}\cdot\textbf{r})|i\rangle|^2
\end{eqnarray}
where a$_0$ is Bohr atomic radius, and vector \textbf{r} is the
coordinate of the electrons in the sample, the initial and final
states of which are represented by $\langle{i}|$ and $\langle f|$
, respectively. Since we have not considered the screening of the
fast electron Coulomb potential by other electrons inside the
material, this expression is only applicable to core electron
excitations.  Because of the inverse q-dependence, electron
scattering is concentrated at small angles. We can then expand the
matrix element in terms of \textbf{q} and only retain the first
non-zero term which is the dipole approximation\cite{EGE96}, to
obtain
\begin{eqnarray}
\frac{d\sigma^2(\textbf{q})}{dEd\Omega}\approx\frac{4}{a_0^2}\cdot\frac{1}{q^4}|\langle
f|\textbf{q}\cdot\textbf{r})|i\rangle|^2
=\frac{4}{a_0^2}\sum_{i,j}\frac{q_iq_j}{q^4}\langle
x_i\rangle\langle x_j\rangle^*\label{b2}
\end{eqnarray}
After projecting \textbf{q} in the sample frame (x,y,z) mentioned
above, one can see the connection between Eq.(\ref{b2}) and
Eq.(\ref{a4}) by identifying $Im(\varepsilon^{ij})\propto\langle
x_i\rangle\langle x_j\rangle^*$. The rest of the derivation can
follow the procedure used in the dielectric formalism.
 Thus, we should arrive at the same conclusion as Eq.(\ref{a7}), so the
magic angle condition should be the same as
Eq.(\ref{macondition}), i.e. $\xi_{\parallel}=\xi_{\perp}$.

\subsection{\label{sec:level2}The solution of the Magic angle
condition} The magic angle condition refers to the convergence and
collection semi-angles, $\alpha_0$ and $\beta_0$ respectively,
which define the experimental set-up where Eq.(\ref{macondition})
is satisfied.  We recall that the fast electron has the simple
energy-momentum relation
$E_0=\displaystyle\frac{\hbar^2k_0^2}{2m}$ so the energy-loss
process must satisfy the following energy and momentum relations
\begin{eqnarray}
E=E-E_0=\displaystyle\frac{\hbar^2(k_0^2-k_f^2)}{2m}
\end{eqnarray}
\begin{eqnarray}
\textbf{q}=\textbf{k}_0-\textbf{k}_f\label{qrelation}
\end{eqnarray}
or
\begin{eqnarray}
q^{2}=k_0^2+k_f^2-2k_ok_f\cos{\theta}\label{qrelation}
\end{eqnarray}

\begin{figure}
\begin{center}
  \includegraphics[scale=0.8]{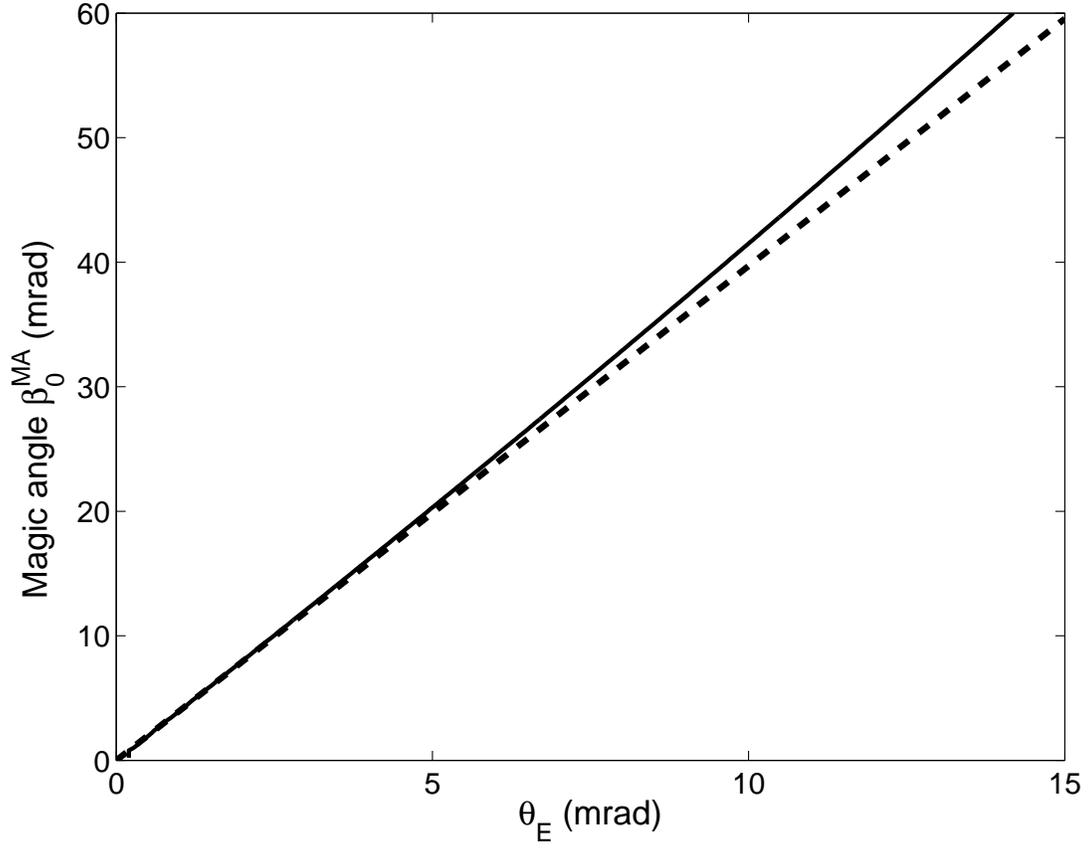}
  \caption{The magic angle is a function of the $\theta_E$ in the parallel illumination case
  under the small angle approximation, i.e. $\beta_0^{MA}$=$3.97\theta_E$(dashed), and for cases no limited by
  the small angle approximation(solid), i.e. solution of Eq.(\ref{macondition}) and (\ref{ap1}).\label{fig3}}
\end{center}
\end{figure}

The simplest test case for magic angle conditions is for an
experimental set-up involving parallel beam illumination.  In this
case, the scattering angle involved ($\theta$) is just the
function of the semi-angle ($\beta$), defined to be the angle
between the wave vector of the scattered electrons and the
electron optical axis.  For an axially placed circular detector,
the maximum and minimum values of the momentum transfer are given
by:
\begin{subequations}
\begin{eqnarray}
&&q_{min}=k_0-k_f=k_0\theta_E\\
&&q_{max}=k_0^2+k_f^2-2k_0k_f\cos{\beta_0}
\end{eqnarray}
\end{subequations}
Using above expressions for q to calculate the integral, following
Paxton et al.\cite{PAX00}, one obtains the result for the reduced
integrals defined in Eq.({\ref{a8}}) as:
\begin{subequations}
\begin{eqnarray}
&&\xi_{\parallel}=A=\frac{1}{8}\frac{2m}{\hbar^2}\frac{E^2}{E_0}(\frac{1}{q_{min}^2}-\frac{1}{q_{max}^2})
+\frac{1}{2}\frac{E}{E_0}\ln{\frac{q_{max}}{q_{min}}}
+\frac{1}{8}\frac{\hbar^2}{2m}\frac{1}{E_0}(q_{max}^2-q_{min}^2)\\
&&\xi_{\perp}=\frac{B-A}{2}=\frac{1}{2}(\ln{\frac{q_{max}}{q_{min}}-{\xi_{\parallel}})}
\end{eqnarray}\label{ap1}
\end{subequations}
This complex solution for the parallel beam illumination can be
simplified because we are interested in the small-angle region
where dipole approximation holds, so we can use the small angle
approximation for q$_{\perp}$ ($\sim k_0\theta$) and
q$_{\parallel}$ ($\sim q_{min}=k_0\theta_E$) as to obtain a more
simplified form for the reduced integrals
\begin{subequations}
\begin{eqnarray}
&&\xi_{\parallel}=A\approx\frac{\hat{\beta}_0^2}{2(\hat{\beta}_0^2+1)}\\
&&\xi_{\perp}=\frac{B-A}{2}\approx\frac{1}{4}[\ln(1+\hat{\beta}_0^2)-\frac{\hat{\beta}_0^2}{\hat{\beta}_0^2+1}]
\end{eqnarray}\label{ap2}\label{parallel}
\end{subequations}
where we have used the reduced collection angle
$\hat{\beta}_0=\beta_0/\theta_E$. We can solve the magic angle
condition using the magic angle relation
$\xi_{\parallel}=\xi_{\perp}$ i.e. $B=3A$.  Within the small angle
approximation, this is satisfied for $\beta_0^{MA}=3.97\theta_E$,
or 4$\theta_E$ approximately as shown initially  for a uniaxial
system\cite{MEN98}. In fig.\ref{fig3}, the magic angle solution
for the parallel beam illumination set-up is plotted as a function
of $\theta_E$. The small-angle solution is found to be acceptable
for the normal energy-loss and collection condition as the
deviation from the more exact solution obtained from
Eq.(\ref{ap1}) and (\ref{ap2}) is less than 5 percent.

\subsection{$\label{sec:level2}$Convergence angle effect}
In many cases, explicitly for a focused probe system or implicitly
because of the need to increase the illumination level at the
sample using a slightly convergent beam, the magic angle solution
for the parallel illumination condition becomes inapplicable. To
take into account the convergence effect, we need to reexamine the
momentum conservation relation shown in Eq.(\ref{qrelation}). This
vector relation can be decomposed according to the vector
components parallel and perpendicular to the electron optical
axis.  In the small-angle approximation (see fig.\ref{fig4}), the
parallel version of this relation is a scala equation:
\begin{eqnarray}
\ q_{\parallel}\equiv -q_Z=k_0\theta_E
\end{eqnarray}
and perpendicular components of \textbf{q} are defined as follows:
\begin{eqnarray}
\left\{
\begin{array}{l}
q_X=k_0(\alpha\cos\phi-\beta\cos\varphi) \\
q_Y=k_0(\alpha\sin\phi-\beta\sin\varphi) \\
q_{\perp}^2=q_X^2+q_Y^2=k_0^2\theta^2\\
\end{array}
\right.
\end{eqnarray}
where $\phi$ and $\varphi$ are azimuth angles for wave-vectors of
the incident and scattered electrons respectively and $\theta$ is
the scattering angle between the direction of the incident
electron and that of the scattered electron. The latter is related
to $\alpha$ ($\beta$), the angle between the incident (scattered)
electron and the electron beam axis, by spherical
trigonometry:(see appendix)
\begin{eqnarray}
\theta^2(\alpha,\beta,\phi-\varphi)=\alpha^2+\beta^2-2\alpha\beta\cos(\phi-\varphi)\label{ab}
\end{eqnarray}

Then, the reduced integral defined in Eq.(\ref{a8}) has the
following form:
\begin{subequations}
\begin{eqnarray}
\xi_{\parallel}=A'
\end{eqnarray}
\begin{eqnarray}
\xi_{\perp}=\frac{B'-A'}{2}
\end{eqnarray}\label{convergence}
\end{subequations}
where
\begin{subequations}
\begin{eqnarray}
A'=\frac{1}{2\pi^2\alpha_0^2}\int_0^{\alpha_0}\alpha d\alpha
\int_0^{2\pi}d\phi\int_0^{\beta_0}\beta d\beta
\int_0^{2\pi}d\varphi\frac{\theta_E^2}{[\theta_E^2+\theta^2(\alpha,\beta,\phi-
\varphi)]^2}
\end{eqnarray}
\begin{eqnarray}
B'=\frac{1}{2\pi^2\alpha_0^2}\int_0^{\alpha_0}\alpha d\alpha
\int_0^{2\pi}d\phi\int_0^{\beta_0}\beta d\beta
\int_0^{2\pi}d\varphi\frac{1}{\theta_E^2+\theta^2(\alpha,\beta,\phi-
\varphi)}
\end{eqnarray}
\end{subequations}
The integral can be evaluated directly as an explicit function of
the convergence and collection angles as:
\begin{eqnarray}
&&A'(\alpha_0,\beta_0,\theta_E)=\frac{1}{4\hat{\alpha_0^2}}(D-C)\nonumber\\
&&B'(\alpha_0,\beta_0,\theta_E)=
\frac{1}{4\hat{\alpha_0^2}}\left\{[2\hat{\beta}_0^2\ln\frac{D+C-2\hat{\alpha_0^2}}{2}
+2\hat{\alpha}_0^2\ln\frac{D +C-2\hat{\beta_0^2}}{2}] -(D-
C)\right\}\nonumber
\end{eqnarray}
where
\begin{eqnarray}
&&C=\sqrt{-4\hat{\alpha}_0^2\hat{\beta}_0^2+(1+\hat{\alpha}_0^2+\hat{\beta}_0^2)^2}\nonumber\\
&&D=\hat{\alpha}_0^2+\hat{\beta}_0^2+1\nonumber\\
&&\hat{\alpha}_0={\alpha_0}/{\theta_E}\nonumber\\
&&\hat{\beta}_0={\beta_0}/{\theta_E}\nonumber
\end{eqnarray}
Thus the solution of Eq.(\ref{macondition}) is equivalent to
$B'=3A'$. Eq.(\ref{convergence}) is equivalent to
Eq.(\ref{parallel}) when the convergence angle $\alpha_0$
approaches zero. The numerical result is plotted as a contour in
fig.\ref{fig5}. Our result agrees with the result given by Souce
et al \cite{SOU98} in their determination for an uniaxial system
through more tedious integration. It is worth pointing out that an
earlier prediction by Menon and Yuan\cite{MEN98} for the magic
angle condition in the convergent beam case is incorrect because
it did not perform the actual azimuthal angular integration for
both the incident beam and the scattered beam.

A striking feature of the solution shown above is the symmetry,
that is, the interchangeability between the beam convergence range
$\alpha_0$ of the incident electron and angular range $\beta_0$ of
the collection of scattered electrons.  This is already evident in
Eq.(\ref{ab}).  This can be traced to the small-angle
approximation. Fig.\ref{fig4} shows that the surface of the Edward
sphere described by the fast electrons becomes a plane of constant
energy in the small angle approximation.  It has been argued that
dependence on $\alpha_0$ and $\beta_0$ are thus not
symmetrical\cite{MEN98}, because the interchange of $\alpha$ and
$\beta$ produces another momentum transfer vector which has the
same parallel vector component but with a perpendicular component
which is pointing in the opposite direction.  Thus in the general
case, the interchangeability of the convergence and the collection
angles may not be valid . But in angular integrated spectroscopy,
and in the small angle approximation, cross-section depends only
on the modular square of \textbf{q}(see Eq.(\ref{a8})), hence the
interchangeability holds.

\section{Discussion of the theoretical model}
Our theoretical analysis not only gives us a definition of the
magic angle conditions, valid for an arbitrary anisotropic system,
but it can also give a general expression for electron energy loss
spectroscopy in anisotropic systems.  This allows us to
investigate in more detail the physical meaning of the magic angle
conditions as well as understanding the discrepancy between the
various reported magic angle analyses.

\subsection{\label{sec:level2}General expression for anisotropic EELS}
A general expression for anisotropic EELS can be worked out by
substituting the result of Eq.(\ref{a10}) back into Eq.(\ref{a4}).
This gives the cross-section for the partially angular integrated
electron energy-loss spectrum for core electron excitation in any
anisotropic material systems in terms of their macroscopic
dielectric function as:
\begin{eqnarray}
\frac{d\sigma}{dE}(\alpha_0,\beta_0,\widetilde{O}) =\frac{8\pi
m_e}{na_0h^2k_0^2}[\xi_{\perp}\emph{\textbf{Im}}[Tr(\varepsilon)]+(\xi_{\parallel}-\xi_{\perp})\sum_{i,j}\cos\chi_{i}\cos\chi_j\varepsilon{_2}^{ij}]
\label{general}
\end{eqnarray}
As the trace of the dielectric function 'metric tensor'
Tr($\varepsilon)$ (=$\displaystyle\sum_j \varepsilon^{jj}$) is
invariant with respect to rotational transformation, the first
part of the expression does not change with specimen orientation.
The second part has a pre-factor that vanishes at the magic angle
conditions. So the cross-section for core electron excitation
using MAEELS is
\begin{eqnarray}
\frac{d\sigma}{dE}(\alpha_0^{MA},\beta_0^{MA}) =\frac{8\pi
m_e}{na_0h^2k_0^2}\xi_{\perp}^{MA}\emph{\textbf{Im}}[Tr(\varepsilon)]
\end{eqnarray}

\subsection{\label{sec:level2}Physical meaning of the magic angle effect}
Another way to write Eq.(\ref{general}) is as follows:
\begin{eqnarray}
\frac{d\sigma}{dE}(\alpha_0,\beta_0,\widetilde{O})=\frac{8\pi
m_e}{na_0h^2k_0^2} \left\{
({\xi_{\parallel}+2\xi_{\perp}})\frac{\emph{\textbf{Im}}[Tr(\varepsilon)]}{3}
+(\xi_{\parallel}-\xi_{\perp})\sum_{i,j}(\cos{\chi_i}\cos\chi_j-\frac{1}{3}\delta_{ij})\varepsilon^{ij}_2
\right\}\label{general2}
\end{eqnarray}
As before, the factor $(\xi_{\parallel}-\xi_{\perp})$ gives the
magic angle condition. However, if we rotate the crystal over all
possible orientations($\widetilde{O}$), the averaged value of the
angular dependent factors can be written as\cite{PRI94}:
\begin{eqnarray}
\overline{(\cos{\chi_i}\cos\chi_j)}=\frac{1}{3}\delta_{ij}
\end{eqnarray}
This means that the second term in Eq.(\ref{general2}) drops out
when the cross section is averaged over all the orientation, even
if the magic angle condition is not satisfied, i.e.
\begin{eqnarray}
\overline{\frac{d\sigma}{dE}}(\alpha_0,\beta_0)=\frac{8\pi
m_e}{na_0h^2k_0^2} \left\{
({\xi_{\parallel}+2\xi_{\perp}})\frac{\emph{\textbf{Im}}[Tr(\varepsilon)]}{3}
\right\}
\end{eqnarray}

Thus we demonstrated that the spectrum obtained at magic angle
condition is equivalent to that obtained by orientational
averaging.  This reason can be explained mathematically as
follows.

In normal orientational averaging, the orientation can refer
either to that of the specimen or the orientation of the external
perturbation, i.e. the orientation of the applied field defined by
the momentum transfer vector \textbf{q} in EELS.  We can
distinguish the averaging over the azimuth angle from 0 to 2$\pi$
, from averaging over the polar angle form 0 to $\pi$.  In MAEELS,
the azimuth angle averaging is achieved by using an axially placed
circular detector. The remaining orientational averaging of
\textbf{q} over the full polar angle range is not possible to
achieve in EELS experiments, but the equivalent result may be
obtained by integrating over a limited range of polar angles
because electron scattering is skewed towards the small angle.
However, it is not always possible to find the appropriate polar
angular range, hence the magic angle condition is not trivial.

\subsection{\label{sec:level2}Comparison with other analysis}
Now we can compare our prediction for the magic angle value with
other derivations (see table.1) and to analysis the reasons for
the diversity of the values predicted.

Menon and Yuan\cite{MEN98} and Souche et al\cite{SOU98} both
derived their values by working out the anisotropic spectral
response in the uniaxial system, but otherwise their conclusions
are identical with ours.  Our general result is in agreement with
their prediction for the specific uniaxial system.

Most interesting is that of Paxton et al. \cite{PAX00} who tried
to derive a general theory for the magic angle condition. In their
paper, the momentum transfer vector \textbf{q} is projected in the
laboratory frame (X, Y, Z) defined above, so we have the quantum
mechanical transitional matrix element as:
\begin{eqnarray}
&&{|\langle f|\textbf{q}\cdot\textbf{r})|i\rangle|^2}=|\langle
f|q_XX+q_YY+q_ZZ|i\rangle|^2 \nonumber\\
&&=q_X^2\langle X\rangle^2+q_Y^2\langle Y\rangle^2+q_Z^2\langle
Z\rangle^2 +2q_Xq_Y\emph{\textbf{Re}}[\langle X\rangle\langle
Y\rangle^*] +2q_Yq_Z\emph{\textbf{Re}}[\langle Y\rangle\langle
Z\rangle^*] +2q_Zq_X\emph{\textbf{Re}}[\langle Z\rangle\langle
X\rangle^*] \label{qm}
\end{eqnarray}
where $\langle X\rangle^2=|\langle f|X|i\rangle|^2$, and $\langle
X\rangle=|\langle f|X|i\rangle|$, the $\langle Y\rangle^2$,
$\langle Z\rangle^2$, $\langle Y\rangle$ and $\langle Z\rangle$
have the similar definition. As discussed above, using the
weighting given in Eq.(\ref{a8}), the angular integrated cross
section can be written as:
\begin{eqnarray}
{\frac{d\sigma}{dE}(\alpha_0,\beta_0,\widetilde{O})\propto}
\quad\xi_{\perp}\cdot(\langle X\rangle^2+\langle
Y\rangle^2)+\xi_{\parallel}\cdot\langle Z\rangle^2
\end{eqnarray}

If $\xi_{\perp}=\xi_{\parallel}=\xi_{0}$,then we have:
\begin{eqnarray}
{\frac{d\sigma}{dE}(\alpha_{0}^{MA},\beta_{0}^{MA})\propto}
\quad\xi_{0}\cdot(\langle X\rangle^2+\langle Y\rangle^2+\langle
Z\rangle^2)
\end{eqnarray}
Paxton et al. \cite{PAX00} derived the magic angle condition by
insisting that the isotropic spectra are given by this equation
without giving detailed explanation. To show that it is
orientation independent, we first consider a case where the
specimen frame coincides with the lab frame, then
\begin{eqnarray}
\langle X\rangle^2+\langle Y\rangle^2+\langle Z\rangle^2=\langle
x\rangle^2+\langle y\rangle^2+\langle z\rangle^2
\end{eqnarray}
As the latter is proportional to the trace of the imaginary part
of the dielectric function, it is invariable to rotation.  So
Paxton et al. \cite{PAX00} chose the correct magic angle
condition, and should arrive at the same value for the magic angle
as we do.  But for reasons we could not understand, they concluded
that the MA$_{\parallel}$ is about 1.36${\theta}$$_{E}$. We
believe that it is a trivial mistake.

Daniels et al.'s derivation\cite{DAN03} has some similarity with
our Eq.(\ref{qm}), but they have used the substitution in the beam
direction coordinate (X,Y,Z) as:
 \begin{eqnarray}
\left\{
\begin{array}{c}
q_{X}=q_{XY}\cos{\phi} \\
q_{Y}=q_{XY}\sin{\phi}
\end{array}
\right.
\\
\left\{
\begin{array}{c}
X=r_{XY}\cos{\phi}'\\
Y=r_{XY}\sin{\phi}'
\end{array}
\right.
\end{eqnarray}
where the r$_{XY}$ was defined as the magnitude of vector
\textbf{r$_{XY}$}-the position vector of the sample electron in
the X-Y plane where the collection aperture lies, and q$_{XY}$ has
a similar definition. However, Daniels et al assumed that
$\phi\neq\phi'$, so their subsequent calculation can not be
correct.

Zhu et al.\cite{ZHU94} correctly recognized the importance of the
rotationally symmetry of the experimental set-up, i.e. that the
cross term as shown in Eq.(\ref{a8}) should vanish in integrating
over azimuthal angle, so they focused on estimating the weighting
of the cross-section along the polar angular range as:
\begin{eqnarray}
\bar{q}_{\parallel}=\bar{q}_{\perp}
\end{eqnarray}
where $\displaystyle \bar{q}_{i}=\int q_i
(\frac{d^2\sigma}{dEd\theta})d\Omega/\int
(\frac{d^2\sigma}{dEd\theta})d\Omega$.  The normalization factor
in the denominator is just the equivalent expression for the
isotropic system and we can define it as N. In the small angle
approximation, we have:
\begin{subequations}
\begin{eqnarray}
&&\bar{q}_{\parallel}=\frac{k_0}{N}\int_o^{2\pi}
d\varphi\int_0^{\beta_0}\theta d\theta
\frac{\theta_E}{(\theta^2+\theta_E^2)}\\
&&\bar{q}_{\perp}=\frac{k_0}{N}\int_o^{2\pi}
d\varphi\int_0^{\beta_0}\theta d\theta
\frac{\theta}{(\theta^2+\theta_E^2)}
\end{eqnarray}
\end{subequations}

In fact $q_{\perp}=k_0 \bar{\theta}$ where $\bar{\theta}$ is the
so called mean scattering angle defined in Egerton's book
\cite{EGE96}, and has a value $\bar{\theta}=2\theta_E
({\hat{\beta}_0-\arctan{\hat{\beta}_0}})/{\ln(\hat{\beta}_0^2+1)}$,
and $q_{\parallel}=k_0{\theta_E}$. i.e. in his definition
$\bar{\theta}=\theta_E$ and we can be resolve
$\beta^{MA}_0\approx1.76\theta_E$ according to this relation.  For
comparison, our equivalent integrals defined in Eq.(\ref{a8}) and
(\ref{macondition}) can be rewritten under the small angle
approximation as:
\begin{subequations}
\begin{eqnarray}
&&\xi_{\parallel}=\int_0^{\beta_0}\theta d\theta
\frac{\theta_E^2}{(\theta^2+\theta_E^2)^2}\\
&&\xi_{\perp}=\frac{1}{2}\int_0^{\beta_0}\theta d\theta
\frac{\theta^2}{(\theta^2+\theta_E^2)^2}
\end{eqnarray}
\end{subequations}

Thus the guess of Zhu et al is incorrect numerically.  By a
similar argument, Gloter et al\cite{GLO00} guessed a different
weighting of transition with \textbf{q} parallel the beam
direction and roughly estimated it with
(\textbf{q}$\cdot$$\textbf{k}$$_0$)$^2$/{q}$^2$$k_0$$^2$. They
also used the isotropic angular distribution as normalized factor.
Remarkably, their guess is correct for parallel illumination case,
but their expression can not correctly account for the convergence
effect because it does not consider the scattering when the
incident and scattered electron beams are not in the same plane as
the beam optical axis.

In summary, we have analyzed the reasons for different prediction
of magic angle values, and found all the discrepancies can be
properly accounted for.  This suggests that there is no
fundamental objections to our theoretical model.

We will discuss the discrepancy between the experimental
measurement\cite{DAN03} and the theoretical predicted value of
magic angle in elsewhere. We want to emphasis here that the
discrepancy is not due to errors in the theoretical analysis, but
rather because of the simplicity of the theoretical assumption or
the experimental interpretation.  We can list a number of factors
that might modify the prediction in our model, such as non-dipole
transition\cite{SAL90}, coherent scattering effect\cite{NEL99},
channelling effect\cite{LEE00}, relativistic
effect\cite{EGE96,FAN56}. But our analysis showed that they may
affect the exact values of the magic angle, but not the conclusion
that magic angle effect, if it exists, applies to all anisotropic
systems and that the spectra collected under magic angle condition
represents an orientation averaging.

\section{Applications}

\subsection{\label{sec:level2}Cross sections in anisotropic systems}
We will discuss special cases for the symmetries exemplified by
crystalline systems, although in fact the systems can be amorphous
or nanostructured.  This allows us to write out the explicit form
of the dielectric function, hence the precise form of the
cross-section for the partially angular integrated EELS core loss
excitation:

i) \emph{Isotropic systems}

This category includes cubic crystals where the non-diagonal
elements vanish and the diagonal elements are identical, of which
the imaginary parts are set to be $\varepsilon_2$.  So the cross
section for EELS is given by:
\begin{eqnarray}
\frac{d\sigma}{dE}=\frac{8\pi m_e}{na_0h^2k_0^2}
\left\{[3\xi_{\perp}+(\xi_{\parallel}-\xi_{\perp})
\sum_i\cos^2\chi_i]\varepsilon_2\right\} \equiv\frac{8\pi
m_e}{na_0h^2k_0^2} (\xi_{\parallel}+2\xi_{\perp})\varepsilon_2
\end{eqnarray} where we have used the relation
$\displaystyle\sum_i\cos^2\chi_i=1$ \cite{PRI94}. According to the
definition of the reduced integral in Eq.(\ref{a8}), the factor
$(\xi_{\parallel}+2\xi_{\perp})$ is in fact proportional to
$\ln(\hat{\beta}_0^2+1)$ in the case of parallel illumination.

ii) \emph{Uniaxial systems}

There are hexagonal, tetragonal and rhombohedral crystals in this
class where the physical response is unique in one direction. The
dielectric function is characterized by two different element as
follows:
\begin{eqnarray}
\varepsilon^{ij}= \left\{
\begin{array}{r@{\quad}l} \varepsilon^{\parallel}& i=j=3\\
\varepsilon^{\perp}& i=j=1or2 \\
0&i\neq j
\end{array}
\right.
\end{eqnarray}
This means that the cross-section for EELS in this class of system
is given by:
\begin{eqnarray}
\frac{d\sigma}{dE}=\frac{8\pi m_e}{na_0h^2k_0^2}\left\{
[\xi_{\perp}(1+\cos^2\chi_3)+\xi_{\parallel}\sin^2\chi_3]\varepsilon_2^{\perp}
+(\xi_{\parallel}\cos^2\chi_3+\xi_{\perp}\sin^2\chi_3)\varepsilon_2^{\parallel}
\right\}\label{uniaxial}
\end{eqnarray}

iii) \emph{Orthorhombic systems}

In this case, the dielectric function have three independent
diagonal elements $\varepsilon^{jj}$, so we have:
\begin{eqnarray}
\frac{d\sigma}{dE}=\frac{8\pi m_e}{na_0h^2k_0^2}\left\{
\sum_j[\xi_{\perp}+(\xi_{\parallel}-\xi_{\perp})\cos^2\chi_j]\varepsilon_2^{jj}
\right\}\label{orthogonal}
\end{eqnarray}

iv) \emph{Monoclinic and Triclinic systems}

The off-diagonal elements do not vanish in these systems, so we
have:
\begin{eqnarray}
\frac{d\sigma}{dE}=\frac{8\pi m_e}{na_0h^2k_0^2}\left\{
\sum_{i,j}[\xi_{\perp}\delta_{ij}+(\xi_{\parallel}-\xi_{\perp})
\cos\chi_i\cos\chi_j]\varepsilon_2^{ij} \right\}
\end{eqnarray}

\subsection{\label{sec:level2}'Magic' orientation in the orthogonal systems}
Here we concentrate on the most commonly encountered orthogonal
system. The cross section Eq.(\ref{orthogonal}) can be rearranged
as:
\begin{eqnarray}
\frac{d\sigma}{dE}=\frac{8\pi
m_e}{na_0h^2k_0^2}\left\{\sum_jW_{jj}\varepsilon^{jj}_2\right\}
=\frac{8\pi
m_e}{na_0h^2k_0^2}\left\{({2\xi_{\perp}+\xi_{\parallel}})\frac{\emph{\textbf{Im}}[Tr(\varepsilon)]}{3}
+(\xi_{\parallel}-\xi_{\perp})\sum_j(\cos^2\chi_j-\frac{1}{3})\varepsilon^{jj}_2\right\}
\end{eqnarray}

Again the first part is rotationally invariant so it represents
the isotropic spectrum.  The second part contains the information
about the specimen rotation.  The first bracket represents the
factor responsible for the magic angle condition.  The interesting
point is the existence of other factors $(\cos^2\chi_j-1/3)$
inside the summation over \emph{j}.  This suggests that the second
part will also vanish if all the brackets within the summation
sign equal to zero.  They uniquely define a specific specimen
orientation ($\chi_j=54.7^{\circ}$) which we may label as the
'magic' orientation.  Again the spectrum so obtained equals that
obtained at the magic angle condition or by through orientational
averaging.  This is understandable as $\cos\chi_j$ is the
projection of the \emph{jth} basis vector at the optical axis, so
each principle symmetry electronic excitation contributes equally.
By rotation symmetry about the beam direction, the same result
holds for the more general case involving convergence
illumination.

In uniaxial systems whose dielectric function has only two
variables $\varepsilon_{\parallel}$ and $\varepsilon_{\perp}$ ,
the angular integrated cross section as shown in
Eq.(\ref{uniaxial}) becomes
\begin{eqnarray}
\frac{d\sigma}{dE} =\frac{8\pi
m_e}{na_0h^2k_0^2}\left\{({2\xi_{\perp}+\xi_{\parallel}})\frac{\varepsilon^{\parallel}_2+2\varepsilon^{\perp}_2}{3}
+(\xi_{\parallel}-\xi_{\perp})(\cos^2\chi_3-\frac{1}{3})(\varepsilon^{\parallel}_2-\varepsilon^{\perp}_2)\right\}
\label{uniax}
\end{eqnarray}

Clearly, the 'magic' orientation defined above is reduced to a
'magic angle' between the z-axis of the sample and the optical
axis, i.e.$\chi_3=54.7^{\circ}$ in uniaxial system.  However, we
have distinguished this 'magic angle' of specimen orientation with
the magic angle for the beam convergence and collection in MAEELS.

In summary, the 'magic' orientation can provide a set up at which
the spectra should be the same as the spectra gained at the magic
angle for the system where the symmetry is higher than orthogonal,
and this 'magic' orientation will lose its meaning in a system
whose dielectric function has the non-zero off-diagonal elements.

Our analysis suggests that a better way to represent the
anisotropic response of EELS is to write it as a linear
combination of the orientationally averaged (also called
'isotropic') spectrum and an orientation dependent spectrum. In
uniaxial systems, the orientation-dependent spectrum can be
further expressed as a product of the magic angle factor, the
magic orientation factor and a dichroic spectrum:
\begin{eqnarray}
\varepsilon_2|_{Anisotropic}=({2\xi_{\perp}+\xi_{\parallel}})\cdot\varepsilon_2|
_{Average}+(\xi_{\parallel}-\xi_{\perp})(\cos^2\chi_3-\frac{1}{3})\cdot\varepsilon_2|_{Dichroic}
\end{eqnarray}
where
\begin{subequations}
\begin{eqnarray}
&&\varepsilon_2|_{Average}=\frac{2\varepsilon^{\perp}_2+\varepsilon^{\parallel}_2}{3}\\
&&\varepsilon_2|_{Dichroic}=\varepsilon^{\parallel}_2-\varepsilon^{\perp}_2
\end{eqnarray}
\end{subequations}

This formula should offer a practical way to study the anisotropy
in the core electron excitation as well as encoding the magic
angle and magic orientation conditions.

\subsection{\label{sec:level2}Connections between MAS for EELS, for NMR and for XAS}
The magic orientation effect has a direct analogue with the 'magic
angle effect' in XAS experiments and less directly with magic
angle spinning nuclear magnetic resonance (MAS-NMR).

In surface extended x-ray absorption fine structure
(SEXAFS)\cite{STH92,KIN88}, one can study the bonding length as
well as local coordination number of the excited atoms. However,
contribution of more than one shell to the measured EXAFS can
result in \emph{a polarization dependent measured distance} from
absorbing atom to the neighboring atoms and affect the effective
coordinate number. If there is higher than twofold symmetry around
the surface normal, the correct distance and the real coordinate
number can be directly measured if the angle, between the electric
field vector \textbf{E} of the incident x-ray and the surface
normal of the single crystal, is equivalent to 54.7$^{\circ}$
exactly.  This is because the system being probed is effectively
an uniaxial system.

Another famous example is the MAS-NMR technique for
solid\cite{RAH86,AND59,LOW59}. If the material, whether it is a
single crystal, polycrystal or power, spun with high speed about
an axis which is respect to the applied magnetic field with
54.7$^{\circ}$ , the NMR result will be independent of the
orientation of the sample. MAS-NMR is also related to the
orientational magic angle effect because one is effectively using
spinning to create an effective 'uniaxial system' out of powered
samples.
\section{Conclusion}
We have presented a general model describing anisotropy of the
core-level electron excitation in EELS measurement and have
determined the magic angle condition at which the sample
orientation becomes irrelevant.  After comparing our derivation
with reported theoretical efforts, we can explain all the reasons
for disagreement in the literature predicting the value of the
magic angle and showed that the differences in no way invalid our
approach. Furthermore, for the first time, we showed that the
magic angle condition is applicable in all anisotropic systems and
that the spectrum at the magic angle condition is equivalent to
the rotational average of the sample. The same analysis can also
give the general expression for electron energy loss spectroscopy
of core electron excitation in anisotropic system.  In high
symmetry cases, it leads to the discovery of the magic orientation
condition. Its relation with other 'magic angle effect' is
clarified.  In addition, the analysis shows that EELS in uniaxial
system can be written as a sum of the effective 'isotropic'
spectrum and the linear dichroic spectrum.

\begin{acknowledgments}
This research is supported by the National Key Research and
Development Project for Basic Research from Ministry of Science
and Technology, the Changjiang Scholar Program of Ministry of
Education and the '100'-talent program of the Tsinghua University.
\end{acknowledgments}

\appendix
\section{the geometrical explanation for Eq.(\ref{ab})}
The perpendicular component of the momentum transfer \textbf{q}
can be seen as a sum of the perpendicular components of the
initial and the final wave-vector in the case of the convergence
beam as:
\begin{eqnarray}
\textbf{q}_{\perp}={\textbf{k}_{0}}^{\perp}-{\textbf{k}_{f}}^{\perp}
\end{eqnarray}

\begin{figure}
\begin{center}
  \includegraphics[scale=0.9]{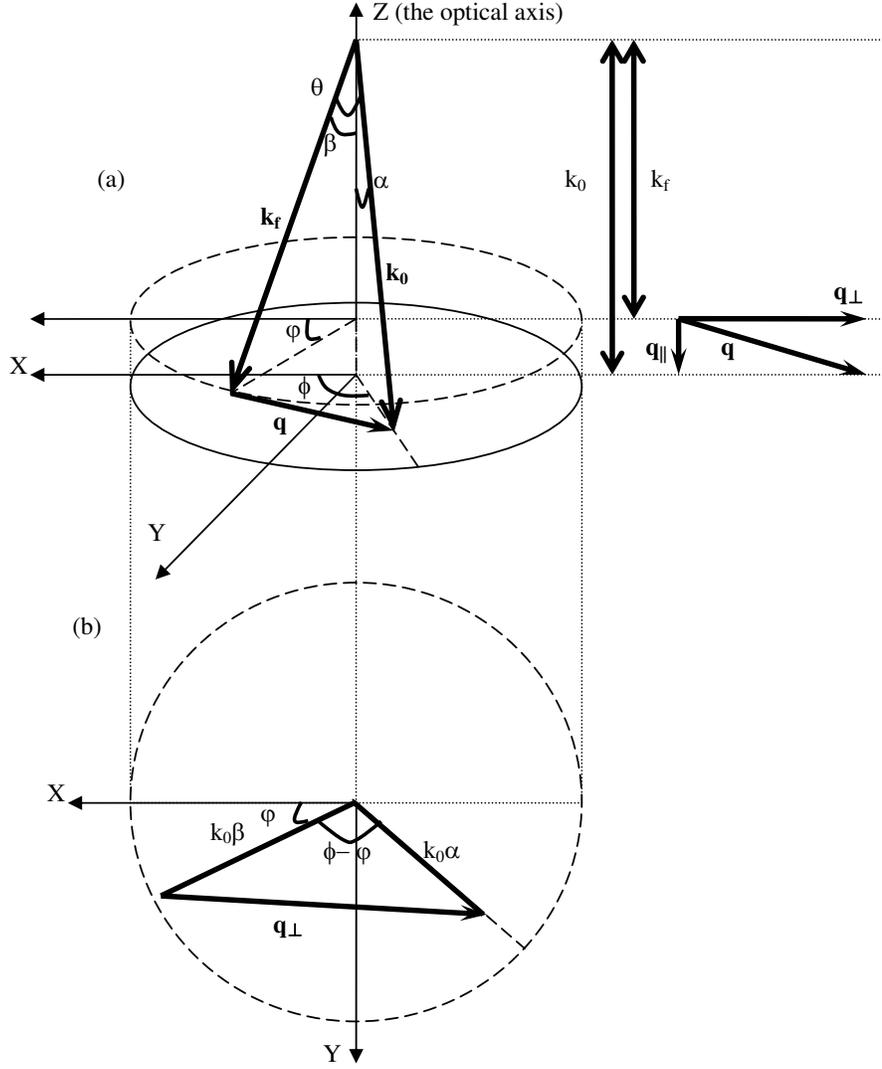}
  \caption{Schematic diagram of the inelastic scattering under the small angle approximation (a) and its projection
  in the plane perpendicular to the optical axis (b).  The convergence angle effect has been taken into consideration.}\label{fig4}
\end{center}
\end{figure}

\begin{figure}
\begin{center}
  \includegraphics[scale=0.7]{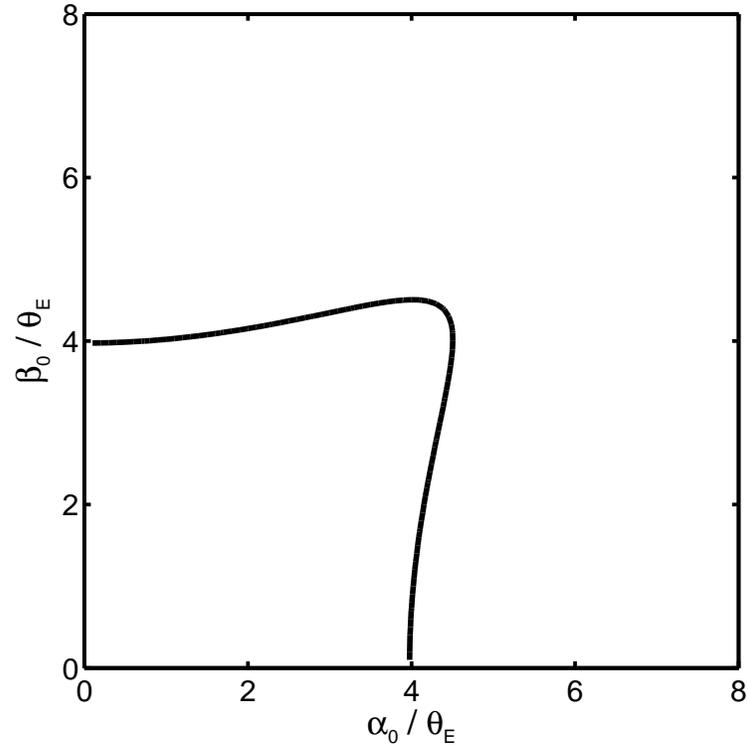}
  \caption{The contour expression for magic angles condition in electron energy loss spectroscopy of anisotropic core electron excitation.}\label{fig5}
\end{center}
\end{figure}

According to the vector relations shown in fig.\ref{fig4} under
the small angle approximation, we have:
\begin{eqnarray}
{q}_{\perp}&=&{k_0}\theta\\
{k_0}^{\perp}&=&{k_0}\alpha\\
{k_f}^{\perp}&=&{k_0}\beta
\end{eqnarray}
Thus the scattering angle $\theta$ can be written following the
vector combination rule as:
\begin{eqnarray}
\vec{\theta}=\vec{\alpha}-\vec{\beta}
\end{eqnarray}
where the directional properties of these angular vectors are
defined to be the same as the perpendicular components of their
corresponding wave-vectors. According to the law of cosines, the
magnitude of $\vec{\theta}$ therefore can be written as:
Eq.(\ref{ab}).
\begin{eqnarray}
\theta(\alpha,\beta,\phi-\varphi)^2=\alpha^2+\beta^2-2\alpha\beta\cos(\phi-\varphi)\nonumber
\end{eqnarray}


\pagebreak
\begin{table}
\caption{\label{table1}The predicted values of the magic
collection angle for the parallel illumination case.}
\begin{ruledtabular}
\begin{tabular}{crrrrr}
  Authors & Zhu\cite{ZHU94} & Menon\cite{MEN98} & Paxton\cite{PAX00} & Souche\cite{SOU98} & Daniels\cite{DAN03} \\
\hline
  MA$_{\parallel}$($\beta_0^{MA}/\theta_E$) & 1.8 & 4 & 1.36 & 3.97 & 1.98
\end{tabular}
\end{ruledtabular}
\end{table}

\end{document}